\documentclass[a4paper,10pt,final]{article}
\usepackage{amsmath,amssymb,amsfonts}
\usepackage{graphics,graphicx}
\usepackage[colorlinks= true,linkcolor=blue,urlcolor=black]{hyperref}
\usepackage[left=4cm,right=3.4cm,top=3.4cm,bottom=3.4cm]{geometry}
\usepackage{natbib}

\title{Formation of conserved charge at the de Sitter space}

\author{V. V. Nikulin\textsuperscript{1}\\
\href{mailto:n-valer@yandex.ru}{n-valer@yandex.ru}
\and S. G. Rubin\textsuperscript{1,2}\\
\href{mailto:sergeirubin@list.ru}{sergeirubin@list.ru}
\and P. M. Petriakova\textsuperscript{1}\\
\href{mailto:petriakovapolina@gmail.com}{petriakovapolina@gmail.com}}

\date{\vspace{15pt}\textsuperscript{1} National Research Nuclear University MEPhI\\(Moscow Engineering Physics Institute)\\115409, Kashirskoe shosse 31, Moscow, Russia\\\vspace{10pt} \textsuperscript{2} Kazan  Federal  University\\420008, Kremlevskaya  street  18, Kazan, Russia}

\begin{document}
\maketitle
\begin{abstract}
The article considers a new mechanism of charge accumulation in the early Universe in theories with compact extra dimensions. The relaxation processes in the extra space metric that take place during its formation lead to the establishment of symmetrical extra space configuration. As a result, the initial accumulation of the number associated with the symmetry occurs. We demonstrate this mechanism using a simple example of a two-dimensional apple-like extra space metric with $U(1)$-symmetry. The conceptual idea of the mechanism can be used to develop a model for the production of the baryon or lepton number in the early Universe.
\end{abstract}

\section{Introduction}
It is known that compact extra dimensions in Kaluza{--}Klein{--}like theories can substantively explain the origin of symmetries observed in particle physics. The main idea of the approach is that the symmetry group of the gauge field theory is considered as a manifestation of the geometric properties of extra space associated with the presence of Killing vectors \cite{Blagojevic}. Nevertheless, there is no answer to the question: “Why is the symmetry group of the standard model exactly the way it is?”. This question simply comes down to the question of the origin of the compact space itself and its geometry.

One of the known problems of the Kaluza{--}Klein theory is the stability of the compact extra space. It can usually be achieved by introducing additional stabilizing fields \cite{2002PhRvD..66b4036C} or by modifying gravity \cite{2006PhRvD..73l4019B}. The stabilization process obviously should take place in an extremely early Universe at the energy scale comparable to the compactification scale. In our previous work \cite{nikulin2019inflationary}, we show how effects caused by the presence of compact extra space can dramatically affect the process of the cosmological inflation and put a limit on the size of the extra dimensions. The present work is a further development of these~ideas.

In this research, we elaborate on the idea of the symmetrization of an extra space metric on the course of the inflationary period. There are no conserved charges from the beginning. As a result of the symmetrization, the metric is endowed with the Killing vector(s) at the end of inflation and the charge is asymptotically conserved. A specific value of the conserved number is caused by the quantum fluctuations during inflation. This idea could be a reason for the baryon asymmetry of the Universe.

%In the process of extra space stabilization at the inflationary stage something very important is actually happening - the appearance of symmetry and conserved numbers.

%%%%%%%%%%%%%%%%%%%%%%%%%%%%%%%%%%%%%%%%%%
\section{Method}

A correct description of how the Universe could arise from the quantum foam can be given only by the theory of quantum gravity that has not yet been developed. Nevertheless, starting with some random configuration below the Planck scale where GR is applicable, we can investigate its further dynamics. In some cases which are discussed in the works \cite{Popov:2019amg}, the three spatial dimensions undergo exponential expansion, while the rest of the spatial dimensions are stabilizing. As the result, the space is divided into a large four-dimensional inflationary Universe and compact extra space.

Despite the fact that we do not understand how the extra space is born, we have no reason to believe that its metric possesses any symmetry since this process is apparently random. However, as a result of further evolution, the extra space undergoes stabilization and symmetrization. The profound reasons for the inevitable emergence of symmetry in this process are associated with the entropy increase as the result of stabilization (the establishment of thermodynamic equilibrium) \cite{Kirillov:2012gy}. Here, we show how such a stabilization process leads to the formation of a nonzero value of the conserved number associated with a symmetry.

\subsection{Accumulation of the Assymetry}\label{sec2.1}

Any symmetries lead to the conservation of certain numbers according to the Noether theorem. When it comes to spatial (extra spatial) symmetries, the conserved numbers can be physically interpreted as the momentum or the angular momentum of fields along the corresponding spatial directions. For instance, it is discussed in~\cite{Blagojevic,cianfrani2005gauge}.

Suppose that such a number associated with the extra spatial symmetry has nonzero value before the beginning of the inflationary period. If the symmetry of the extra space metric is not broken at this stage, the total number is conserved and its density is quickly diluted in the course of inflation. However, if the symmetrization and stabilization of the metric continue during the inflationary period, this number can be generated and could have non{--}zero value after the end of the inflation. The final relaxation of the extra space metric leads to the appearance of symmetry and, consequently, to the conservation of the accumulated number.

{Let the Killing vector field $\xi^a(x)$ characterize the extra{-}spatial symmetry. In this case, the Lie derivative of the extra metric along the Killing vector field $L_{ \xi}g_{d,mn}=0$ and the extra{--}spatial metric $g_{d,mn}$ remains invariant under the infinitesimal transformations $x_m\rightarrow x_m+\xi^m(x)$. According to the Noether’s theorem (see technical details in \cite{Blagojevic}), there is a conserved current associated with such invariance $\partial_a J^a=0$. This current for any physical field $\chi$ is given by}
\begin{equation}
J^a=\frac{\partial L_\text{m}(\chi)}{\partial(\partial_a \chi)}\xi^b\partial_b\chi - \xi^a L_\text{m}(\chi)\,
\end{equation}
where $L_m$ is a Lagrangian of matter.
The corresponding conserved number
\begin{equation}\label{numb}
Q (t) =\int J^0 \sqrt{|g_4|} \sqrt{|g_d|}\, d^3x\,d^dy\,
\end{equation}
has the meaning of extra spatial (angular) momentum.
Until the metric acquires a stable configuration, this number is not conserved, but its accumulation over time can be calculated.
It is necessary to simulate the metric and scalar field variation in the course of stabilization in order to calculate this accumulated number. The next subsection is devoted to this subject.

\subsection{Extra Spatial Dynamics}

As an example, consider a compact two{-}dimensional apple{-}like extra space formed as a final stage of the stabilization. The stationarity of this configuration was shown in the works \cite{2017JCAP...10..001B,rubin2020cosmology}. In addition, it has axial symmetry which can be interpreted (in the effective four{-}dimensional theory) as $U(1)${-}symmetry with the corresponding conserved number. In contrast to the simplest case of the $U(1)${-}symmetry of a one-dimensional circular extra space (see for example in \cite{sarkar2007particle}), our two-dimensional apple-like extra space may lose its $U(1)${-}symmetry due to the metric perturbations caused by internal gravitational~dynamics.

The  modified $f(R)${-}gravity is used in the following discussion to stabilize such a configuration, even in the absence of a scalar field. The parameters of a four-dimensional $f(R)${-}gravity are limited by a variety of modern observations; see, for example, \cite{berry2011linearized}. However, parameters of higher-dimensional versions of $f(R)${-}gravity can be varied in a wide range since its connection with observational data are not~direct.

Firstly, let us determine the stationary configuration in the $f(R)${--}model. The action is taken in the~form
\begin{eqnarray}\label{act1}
S = \frac{m_D ^{D-2}}{2}\int d^{D}Z \sqrt{|G|}\left[f(R)+L_\text{m}\right]\,, \qquad f(R) = aR^2 + R+c\,.
 \end{eqnarray} 
Here, $D=d+4$, $m_D$ is higher dimentional Planck mass and $L_m$ is a Lagrangian of matter. A charge is accumulated by fields during the stabilization. We will investigate the simplest case of a massive scalar field:
\begin{eqnarray}
L_\text{m}=\frac{1}{2} G^{MN}\partial_M\chi \partial_N\chi-V(\chi)\,,\qquad
V(\chi)=\frac{1}{2}m^2\chi^2\,.
\end{eqnarray}

Consider a Riemannian manifold with metric
\begin{equation}\label{metric}
ds^2 = G_{MN}dZ^M dZ^N = g_{\mu\nu}(x)dx^{\mu}dx^{\nu} + g_{d,mn}(x,y)dy^m dy^n. \, 
\end{equation} 
Here, $M_4$, $K$ are the manifolds with metrics $g_{\mu\nu}(x)$ and $g_{d,mn}(x,y)$, respectively.  %$x$ and $y$ are the coordinates of the subspaces $M_4$ and $K$.
We will refer to four{-}dimensional space $M_4$ and $d${--}dim compact space $K$ as a main space and an extra space respectively. The metric has the signature (+ - - - ...), the Greek indices $\mu, \nu =0,1,2,3$ refer to four{-}dimensional coordinates. Latin indices run over $m,n = 4, ..., d+3$ and refer to extra space. Throughout this paper, we use the conventions for the curvature tensor $R_{ABC}^D=\partial_C\Gamma_{AB}^D-\partial_B\Gamma_{AC}^D+\Gamma_{EC}^D\Gamma_{BA}^E-\Gamma_{EB}^D\Gamma_{AC}^E$ and for the Ricci tensor $R_{MN}=R^A_{MAN}$. The units $\hbar = c = 1$ are also used.

A time behavior of the metric tensor $G_{MN}(x,y)$ is governed by the classical equations of motion and changes under variations of initial conditions. As was shown in \cite{Kirillov:2012gy}, the energy dissipation into the main space $M_4$ leads to the entropy decrease of the manifold $K$. This explains an emergence of a friction term in the classical equations for the extra metric $g_{d,mn}(x,y)$ which stabilizes the extra metric. Finally, the inflationary process strongly smooths out space inhomogeneity so that
\begin{equation}\label{uniform}
g_{d,mn}(x,y)
\xrightarrow{t\rightarrow\infty}g_{d,mn}(t,y)
\end{equation}
at the present epoch. Time dependence of the external metric was discussed within the framework of the Kaluza{--}Klein cosmology and Einstein's gravity \cite{Abbott:1984ba}. If a gravitational Lagrangian contains terms nonlinear in the Ricci scalar, the extra metric $g_{d,mn}$ could have asymptotically stationary states \cite{2006PhRvD..73l4019B,Kirillov:2012gy}
\begin{equation}\label{statio}
g_{d,mn}(t,y)\rightarrow g_{d,mn}(y).
\end{equation}
For additional information, see \cite{2002PhRvD..66b4036C,2002PhRvD..66d5029N}.

To simplify the problem, we assume that $g_{\mu\nu}$ is the metric of the de-Sitter space
\begin{equation}\label{H}
g_{\mu\nu} = diag(1,-e^{2Ht},-e^{2Ht},-e^{2Ht})\, ,
\end{equation}  
where $H$ is the Hubble parameter. 
The four{-}dimensional metric is caused by the inflaton field; the dynamics of which is not considered here.

To find the stationary configurations of metric, we will use the Einstein equations for $f(R)${-}gravity:
\begin{equation}\label{eqn}
R_{MN} f' -\frac{1}{2}f(R)g_{d,MN} 
+ \nabla_M\nabla_N f' - g_{d,MN} \square f' = \frac{1}{m_D^{D-2}}T_{MN}.
\end{equation}
Here, $\square$ stands for the d'Alembert operator
\begin{equation}\label{dalamb}
\square =\frac{1}{\sqrt{|G|}}\partial_M ( G^{MN}\sqrt{|G|}\partial_N)\,.
\end{equation}
It is assumed that the Einstein equations related to the dynamics of four{-}dimensional space are satisfied provided that  the four-dimensional metric $g_{\mu\nu}$ \eqref{H} is postulated.

The stress{--}energy tensor $T_{MN}$ is defined as
\begin{equation}\label{TEI}
T_{MN} = -2 \frac{\partial L_\text{m}}{\partial G^{MN}} + G_{MN} L_\text{m} \, .
\end{equation}
Since the uniform three{-}dimensional dynamics of the inflationary Universe is fixed by hand, we will assume that the scalar field is constant in three{-}dimensional space and varies only in the extra space. For a scalar field $\chi$, we have:
\begin{equation}\label{mout}
\square_d\chi = - V'(\chi) \, ,
\end{equation}
where $\square_d$ is an extra dimensional d'Alembert operator. 

At the end of stabilization process, the relaxation of the extra space can be considered as the damping of small perturbations of metric on a stationary configuration.

Further strategy is as follows. We will find the metric and field of stationary configuration of the extra space by solving the system of Equations \eqref{eqn} and \eqref{mout}. Then, we study the time evolution of small fluctuations near the stationary configuration by solving the linearized system equations. This information will allow for calculating the charge \eqref{numb} accumulation with time.

%%%%%%%%%%%%%%%%%%%%%%%%%%%%%%%%%%%%%%%%%%
\section{Results}

\subsection{Stationary Configuration}
We assume that two{-}dimensional sphere{-}like extra space metric \eqref{uniform} has the form
\begin{equation}\label{2dmetr}
g_{2,mn}=\left(\begin{array}{cc}
-r^2e^{2\beta(t,\theta,\phi)}&0\\
0&-r^2e^{2\beta(t,\theta,\phi)}\sin^2{\theta}\\
\end{array}\right)\,.
\end{equation}

Firstly, let us find the stationary, axially symmetric configuration $\beta(t,\theta,\phi)=\beta_\text{st}(\theta)$ and for the scalar field the same $\chi(t,\theta,\phi)=\chi_\text{st}(\theta)$.
This metric has the Killing vector along $\phi$ coordinate (rotational symmetry) so that the corresponding conserved number exists. Such metrics were used in \cite{2017JCAP...10..001B}.

%As the equations of the system, we will use the trace (excluding three{-}dimensional spatial components) from the Einstein equations \eqref{eqn}, the relation between the metric function $\beta_\text{st}(\theta)$ and the Ricci scalar $R_\text{st}(\theta)$ (to reduce the order of the system to 2) and equation of motion for scalar field \eqref{mout}. 

The system of three equations for three unknown functions $R(\theta)$, $\beta(\theta)$, $\chi(\theta)$ can be obtained from the Einstein Equation  \eqref{eqn}
\begin{equation}\label{s0}
\begin{cases}
 e^{2 \beta } r^2 \Bigl(3 \bigl(a R^2+c+R\bigr)-6 H^2 \bigl(2 a R+1\bigr)-3 \kappa  m^2 \chi ^2\Bigr)+4 \bigl(2 a R+1 \bigr) \beta _{\theta } \cot \theta + \\
 + 4 \bigl(2 a R+1 \bigr) \beta _{\theta \theta } - 8 a \bigl(\cot \theta  R_{\theta }+R_{\theta \theta }\bigr)+4 \bigl(2 a R+1\bigr)+\kappa  \chi _{\theta }^2 =0 \, ,  \\
e^{2 \beta } r^2 \bigl(12 H^2-R\bigr)-2 \bigl(\beta _{\theta } \cot \theta +\beta _{\theta \theta }-1\bigr)=0 \, ,  \\
e^{2 \beta } m^2 r^2 \chi - \chi _{\theta } \cot \theta +\chi _{\theta \theta }=0 \, , 
\end{cases}
\end{equation}
where $\kappa=1/m_6^4$ is the inverse 6{--}dim Plank mass. The second equation follows from the definition of the Ricci scalar.

Numerical calculations need specific values of the parameters and additional conditions. Let the model parameters be $r = 2$, $H = 0.03$, $a = -1$, $c = -1$, $m = 1$, $\kappa = 1$ and the self{--}consistent boundary conditions at the point $\theta=0$:
\begin{align}\label{BC}
\beta(0) &=0\,,\qquad\ \ \, \beta_\theta(0)= 0\,,\nonumber\\
R(0) &=\frac{2}{r^2}\,,\qquad R_\theta(0)=0\,,\\
\chi(0) &=0.1\,,\qquad \chi_\theta(0)=0\,.\nonumber
\end{align}

\begin{figure}[h]
\center{\includegraphics[width=0.9\textwidth]{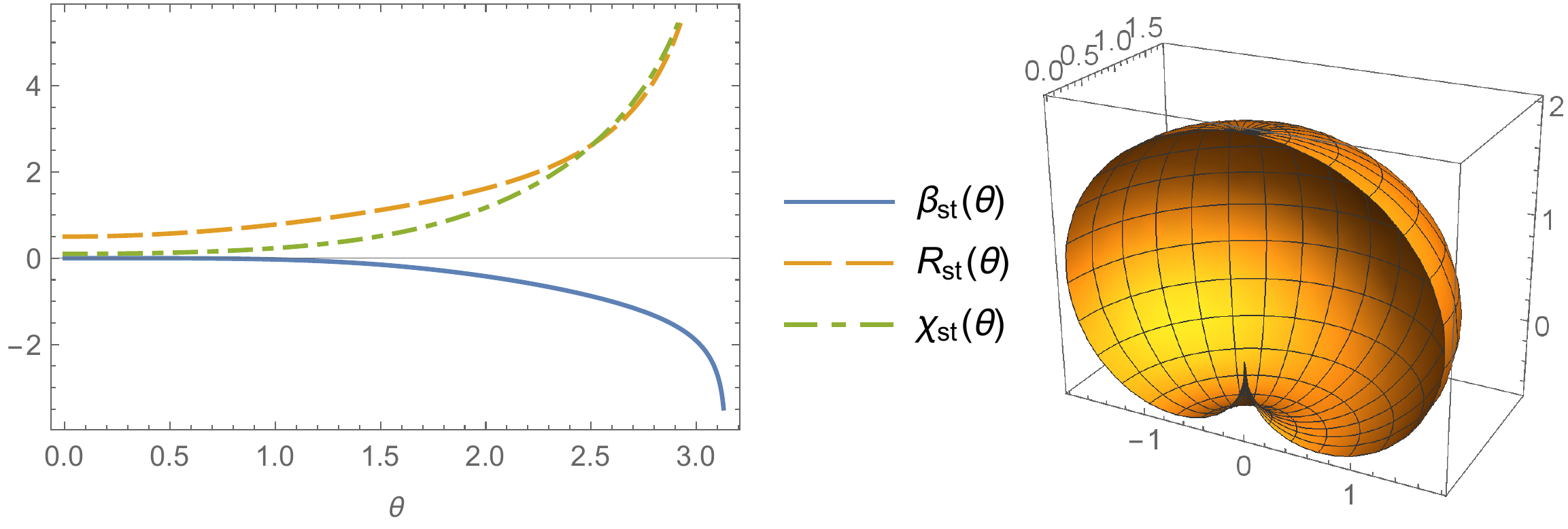}} 
\caption{On the left: dependence of the stationary metric parameter $\beta_{\text{st}}$, the Ricci scalar $R_{\text{st}}$ and the scalar field $\chi_{\text{st}}$ on the extra space azimuthal angle $\theta$. On the right: image of the shape of two-dimensional extra space defined by the metric \eqref{2dmetr}.}
\label{Stationary}
\end{figure}

The solution to the system \eqref{s0} with chosen parameter values and boundary conditions \eqref{BC} is shown in Figure \ref{Stationary}.

\subsection{Small Perturbations}

As the next step, consider small perturbations to the static metric, Ricci scalar and scalar field in the following form:
\begin{align}\label{dchi}
\beta(t,\theta,\phi) &=\beta_\text{st}(\theta)\,+\delta\beta(t,\theta,\phi)
\,\qquad \, \delta\beta(t,\theta,\phi) \ll \beta_\text{st}(\theta) \, , \nonumber\\
R(t,\theta,\phi) &=R_\text{st}(\theta)+\delta R(t,\theta,\phi)
\,\qquad \delta R(t,\theta,\phi) \ll R_\text{st}(\theta) , \\
\chi(t,\theta,\phi) &=\chi_\text{st}(\theta)\, +\delta\chi(t,\theta,\phi)
\,\qquad \, \delta\chi(t,\theta,\phi) \ll \chi_\text{st}(\theta) \, . \nonumber
\end{align}
To study the dynamics of small perturbations, we linearize the system of equations similar to those obtained in the previous paragraph but with respect to all variables:
\begin{equation}\label{syst}
\begin{cases}  
e^{2 \beta }r^2\Bigl(3 \delta R \bigl(4 a H^2-2 a R-1\bigr) + 8 a \delta R_{\phi \phi } \csc ^2 \theta - 8 a \delta R \bigl(\beta _{\theta } \cot \theta +\beta _{\theta \theta }-1\bigr)+ \\ + 2\Bigl(3 \delta\chi \kappa  m^2 \chi -3 \delta\beta \Bigl(R \bigl(-4 a H^2+a R+1\bigr)+c-2 H^2-\kappa  m^2 \chi ^2\Bigr) - \\ - 12 a H R \delta\beta_{t} + 4 \bigl(2 a R+1 \bigr) \delta\beta_{tt} +18 a H \delta R_{t}-4 a \delta R_{tt}+6 H \delta\beta_{t}\Bigr)\Bigr) + \\ + 8 a \delta R_{\theta } \cot \theta - 4 \delta\beta_{\theta \theta } - 4 \bigl(2 a R+1 \bigr) \delta\beta_{\phi \phi } \csc ^2 \theta -  4 \bigl(2 a R+1\bigr) \delta\beta_{\theta } \cot \theta + \\ + 8 a \bigl(\delta R_{\theta \theta }- R \delta\beta_{\theta \theta }\bigr) + 2\kappa  \delta\chi_{\theta } \chi _{\theta }=0 \, , \\ \\
e^{2 \beta } r^2 \Bigl(\delta \beta \bigl(24 H^2-2 R\bigr)+4 \bigl(3 H \delta \beta_{t}+\delta \beta_{tt}\bigr)-\delta R\Bigr)-\\-2 \Bigl(\delta \beta_{\theta } \cot \theta +\delta \beta_{\phi \phi } \csc ^2 \theta +\delta \beta_{\theta \theta }\Bigr) =0 \, , \\ \\
e^{2 \beta } r^2 \Bigl(m^2 \bigl(2 \delta \beta \chi +\delta \chi \bigr)+3 H \delta \chi_{t}+\delta \chi_{tt}\Bigr)-\delta \chi_{\theta } \cot \theta -\delta \chi_{\phi \phi } \csc ^2 \theta -\delta \chi_{\theta \theta }=0 \, .
\end{cases}
\end{equation}
To simplify the solutions, we expand them in the following form:
\begin{align}\label{m}
\delta\beta(t,\theta,\phi) & = \textstyle\sum\limits_{n} \delta\beta_n(t,\theta) \sin(n\phi)\,, \nonumber\\ 
\delta R(t,\theta,\phi) & = \textstyle\sum\limits_{n} \delta R_n(t,\theta) \sin(n\phi)\,, \\
\delta\chi(t,\theta,\phi) & = \textstyle\sum\limits_{n} \delta\chi_n(t,\theta) \sin(n\phi)\,.\nonumber
\end{align}
In this case, the system of Equation \eqref{syst} can be solved separately for each mode.
%\begin{eqnarray}
%\delta\beta(t,\theta,\phi)&=& \sum\limits_{n} \delta\beta_n(t,\theta) \sin(n\phi)\,\nonumber\\
%\delta R(t,\theta,\phi)&=& \sum\limits_{n} \delta R_n(t,\theta) \sin(n\phi)\,\nonumber\\
%\delta\chi(t,\theta,\phi)&=& \sum\limits_{n} \delta\chi_n(t,\theta) \sin(n\phi)\,.
%\end{eqnarray}
Since we consider the final stages of the relaxation process of the extra space, we are not interested in high{-}energy modes which dissipate their energy first. Let us consider the $n=2$ mode.
We choose the simplest initial and boundary conditions to demonstrate how the Ricci scalar oscillations affect the field dynamic. %Thus, the only nonzero initial perturbation are chosen to be the perturbation of $\delta R (0,\theta)$:
\begin{align}
\delta\beta(0,\theta)&=\delta\beta_t(0,\theta)=\delta\beta_\theta(t,0)=\delta\beta_\theta(t,\pi)=0 \, , \nonumber \\
\delta R(0,\theta)&=0.1 \sin^2 \theta\,\,\,\,\,\delta R_t(0,\theta)=\delta R_\theta(t,0)=\delta R_\theta(t,\pi)=0 \, , \\
\delta\chi(0,\theta)&=\delta\chi_t(0,\theta)=\delta\chi_\theta(t,0)=\delta\chi_\theta(t,\pi)=0 \, \nonumber .
\end{align}%\right.

\begin{figure}[h]
\center{\includegraphics[width=0.8\textwidth]{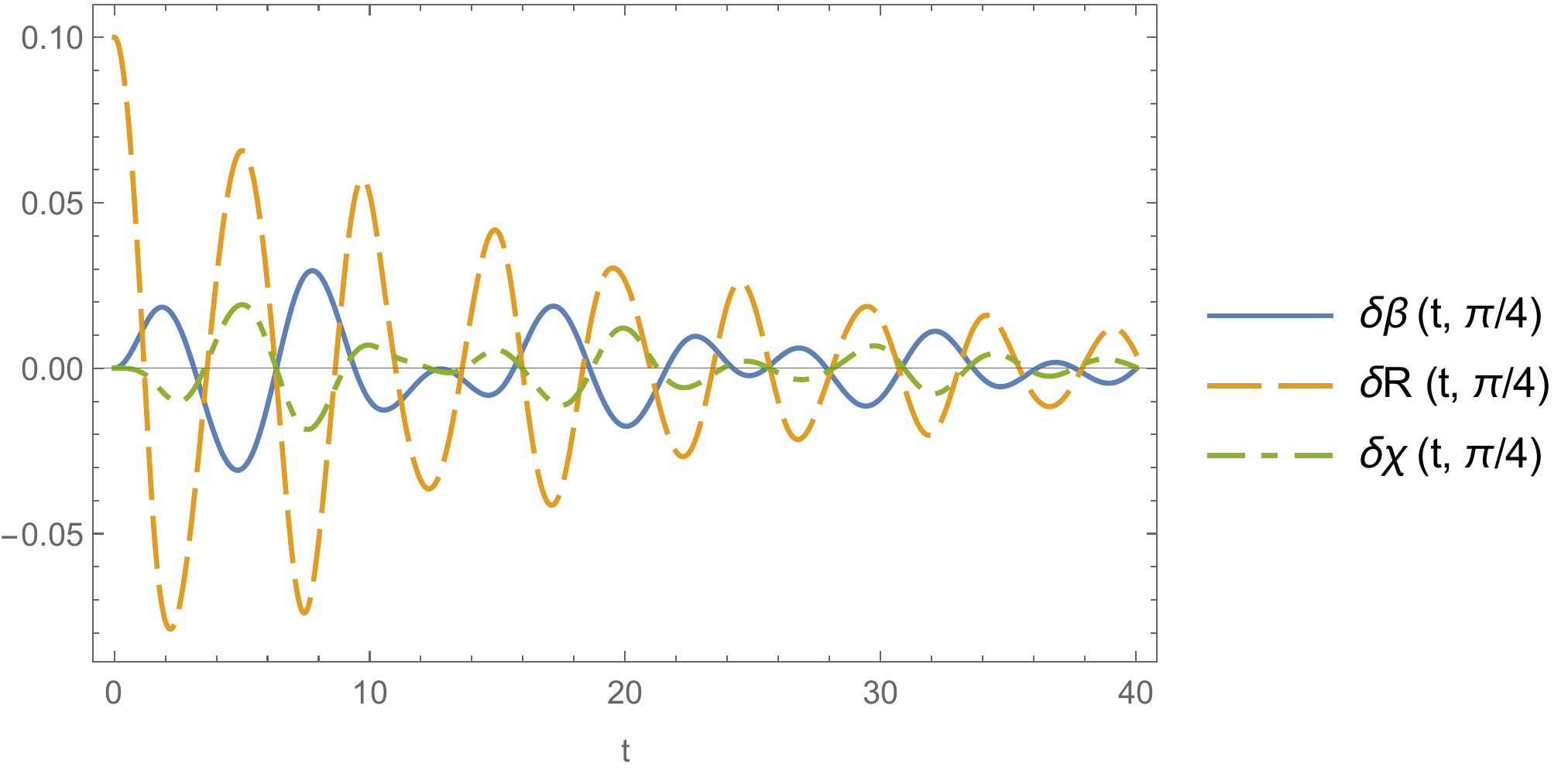}}
\caption{Time evolition of perturbations of the metric parameter $\delta\beta$, the Ricci scalar $\delta R$ and the scalar field $\delta\chi$ for angle $\theta=\pi/4$. The perturbation are the $n=2$ standing wave mode along $\phi$ coordinate.}
\label{Osc}
\end{figure}

The result of calculations are presented in Figure \ref{Osc}. One can see the gradual damping of the oscillations of all sorts. The dumping similar to those shown in Figure  \ref{Osc} occurs for all angles $\theta$. The Hubble friction term leads to the damping of perturbations and stabilization to a stationary~configuration.

\subsection{Accumulation of the Conserved Number}

{According to Section \ref{sec2.1}, the number defined by Formula \eqref{numb} must be conserved due to $U(1)$-symmetry. However, we are interested in how the number $Q$ could be initially accumulated and then conserved. The process of accumulation occurs through a violation of Noether’s theorem by a perturbation of the symmetric state of the metric. The perturbations of the metric and scalar field $\delta \beta, \delta R,\delta \chi$ are related to each other by linear system of Equation  \eqref{syst}. The solutions obtained in the previous subsection allow us to calculate the number $Q$. We are interested in the three-dimensional density of the number in the accompanying volume. The invariant three-dimensional integration in Formula \eqref{numb} is discarded and, after substitution of solutions for the metric and scalar field, we get

%As we already mentioned in the subsection 2.1 that for $U(1)$-symmetry reasons, a certain number  \eqref{numb} must be conserved.The solutions obtained in the previous subsection allow us to calculate it. We are interested in the three{-}dimensional density of the number in the accompanying volume. The invariant three{-}dimensional integration in \eqref{numb} is discarded and after substitution of field and metric solutions we get it in the following form:
\begin{align}\label{numb2}
Q(t)&=\int \partial^0 \chi \partial_\phi \chi\, r^2 e^{2\beta} \sin \theta\, d\theta d\phi = \\
&= \int \partial^0 \delta\chi(t,\theta,\phi) \partial_\phi \delta\chi(t,\theta,\phi)\, r^2 e^{2\bigl(\beta_{st}(\theta)+ \delta\beta(t,\theta,\phi)\bigr)} \sin \theta\, d\theta d\phi \, . \nonumber
\end{align}}
%The metric and field perturbations $\delta \beta, \delta R,\delta \chi$ are related to each other by linear system of equations \eqref{syst}. Thus, 

{If we have stationary metric $\beta(t,\theta,\phi) = \beta_{st}(\theta)$ with axial symmetry, then perturbation of a scalar field $\delta \chi$ is the solution of the equation of motion \eqref{mout} with nonperturbed d'Alambert operator. %For such a solution and $\delta\beta =0$
In this case, Noether's theorem states that the number \eqref{numb2}, which directly follows from Formula \eqref{numb}, will be conserved. The physical meaning of the quantity $Q(t)$ is an internal angular momentum along $\phi${-}coordinate and the conservation of this quantity looks natural due to the rotational symmentry of stationary extra space.} %The state mentioned above is post-inflationary. However, we are interested in how the number $Q$ that conserves after inflation, could be initially accumulated.
%This process occurs through a violation of Noether's theorem by a perturbation of the symmetric state of the metric. 

{We are interested, as was already mentioned, in how the number $Q$, which is conserved after inflation, could be initially accumulated. Therefore, we use perturbed solutions of system of Equation \eqref{syst} for numerical calculation of the time dependence of $Q(t)$. As a perturbation, we take traveling wave superposition with $n = 2$}
\begin{align}
\delta\beta(t,\theta,\phi) & =  \delta\beta_2(t,\theta) \sin(2\phi)+\delta\beta_2(t + \frac{n \pi}{2},\theta) \cos(2\phi)\,, \\
\delta\chi(t,\theta,\phi) & =  \delta\chi_2(t,\theta) \sin(2\phi)+\delta\chi_2(t + \frac{n \pi}{2},\theta) \cos(2\phi)\,.\nonumber
\end{align}

\begin{figure}[h]
\center{\includegraphics[width=0.7\textwidth]{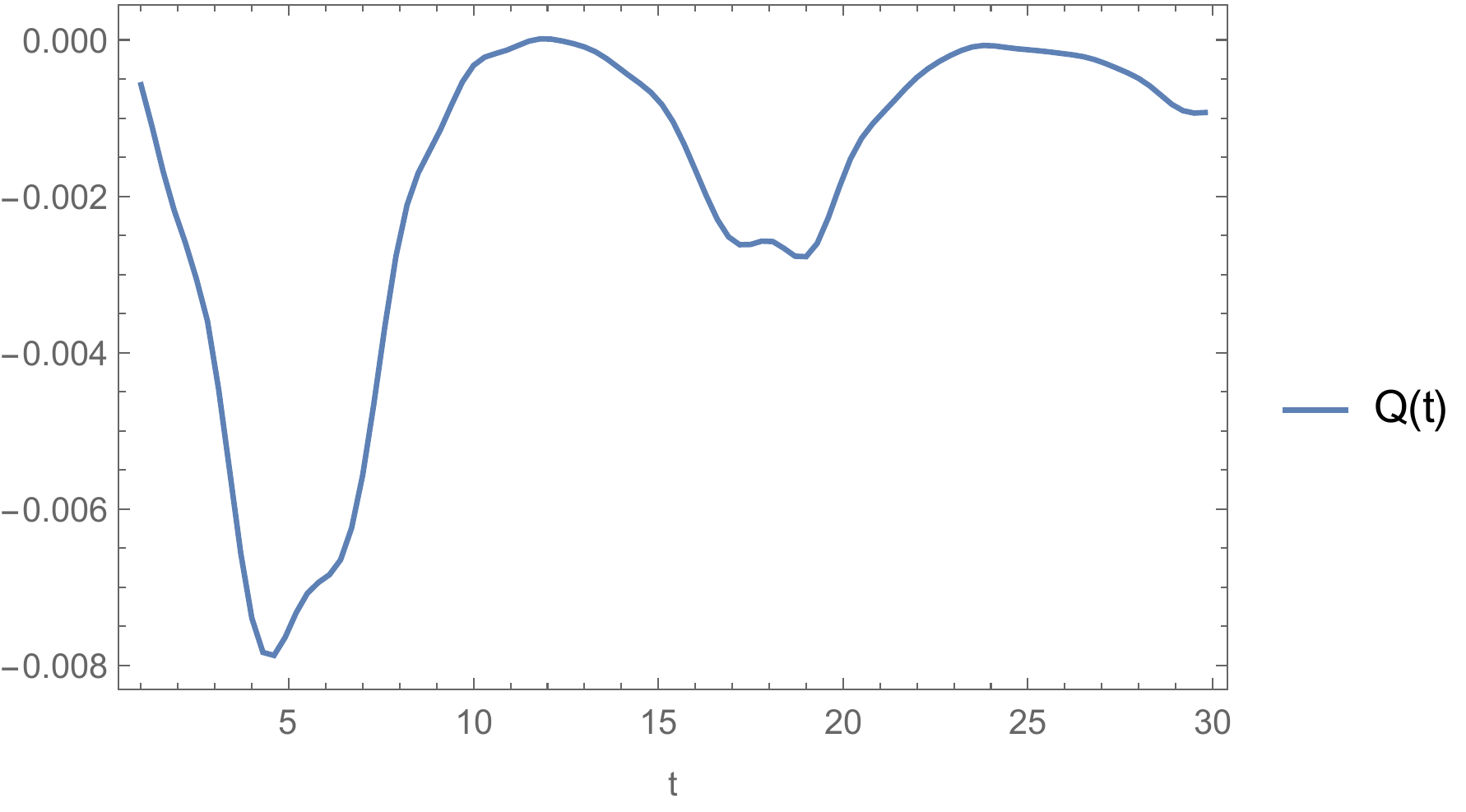}} 
\caption{Time evolution of the symmetry associated number $Q$ during the relaxation of extra space~metric.}
\label{Charge}
\end{figure}

Thus, a traveling wave in the metric generates a traveling wave in a scalar field $\chi$ leading to nonconservation of the number $Q$, which is shown in Figure \ref{Charge}. {The end of inflation is characterized by a relatively sharp transition associated with a violation of the slow-roll conditions. After such a transition, the metric is symmetrized (extra metric fluctuations are quickly suppressed when $H\lesssim 1/r$), while the fields go into a rapidly oscillating mode and the stage of reheating begins. Field fluctuations (traveling waves) generated in extra space are now enclosed there forever, since the internal moment $Q$ is now conserved.} The number $Q(t)$ accumulated by that time will remain unchanged in the future. The Universe enters the stage of the hot Big Bang with a nonzero initial value of this number and all the ensuing consequences.

%%%%%%%%%%%%%%%%%%%%%%%%%%%%%%%%%%%%%%%%%%
\section{Conclusion and Further Research}

In this work, it was shown how the dynamics of the extra space metric lead to the accumulation of a conserved charge. Such dynamics should naturally occur in the early multidimensional Universe at the stage of separation and stabilization of the extra space to a stationary configuration, the symmetry of which is associated with a conserved (and accumulated at an early stage) charge.

%Our task was to show the idea of the mechanism in almost pure form using the simplest example of a scalar field in a two{-}dimensionalensional apple{--}like extra space.

A similar mechanism of the conserved number accumulation can be used in perspective to explain the phenomenon of the baryon asymmetry \cite{2006PZETF...83..3,2009NuPhB.807..229D} though other mechanisms also exist \cite{Belotsky_2014}. The baryon number is known to be associated with the global $U(1)${-}symmetry. The latter could be realized as the axial symmetry of the two{-}dimensional extra space (with metric \eqref{2dmetr}). However, the additional mechanism to transfer the baryon charge from a scalar field to the spinor fields is required, see, e.g., \cite{1993PhRvD..47.4244D} for details. 

{Briefly, the mechanism is realized by the presence of the Yukawa coupling between the multidimensional spinor and scalar fields
\begin{align}
S_{\text{Y}}&\sim\int dy\,\chi(x,y)\bar{\Psi}(x,y)\Psi(x,y) = \\ &= \int dy\,\phi_n(x)\bar{\psi}_l(x)\psi_k(x)e^{i(n-l-k)y} = \delta_{n,\,l+k}\phi_n\bar{\psi}_l\psi_k. \nonumber
\end{align}
Here, only the coordinate $y$ for which the metric possesses symmetry is written explicitly. Such term is known that contains the conservation of the energy and momentum. Therefore, if the scalar particle decays into other particles, the momentum along the $y$ coordinate should be conserved. The equality $n=l-k$ means that, during $\phi\rightarrow\bar{\psi}\psi$ decay, the total moment of the scalar field $n$ goes into the spinor field $l+k$ and is conserved. Numbers $l$ and $k$ are associated with the number of spinor particles and antiparticles, and a nonzero initial $n$ is converted to a nonzero difference in the number of particles and antiparticles.}
%After the integration over the extra dimensions, an effective four-dimensional coupling between the spinor and scalar levels of the Kaluza-Klein tower appears. 
%If there is an initial $U(1)${--}asymmetry in a scalar field, it will be transferred to the spinor field in the form of an imbalance of particles and antiparticles. 

 {However, we foresee difficulties associated with the mass of particles and the Higgs mechanism in this case. Discussion on this subject and a detailed implementation of the mechanism is planned for the future studies.}

%%%%%%%%%%%%%%%%%%%%%%%%%%%%%%%%%%%%%%%%%%
\vspace{6pt} 

%%%%%%%%%%%%%%%%%%%%%%%%%%%%%%%%%%%%%%%%%%
%% optional
%\supplementary{The following are available online at \linksupplementary{s1}, Figure S1: title, Table S1: title, Video S1: title.}


\section{Acknowledgments}

The authors are grateful to Konstantin M. Belotsky and Maxim Yu. Khlopov for helpful discussions.
The work of S.G.R. is supported by the grant RFBR N~19-02-00930 and is performed according to the Russian Government Program of Competitive Growth of Kazan Federal University. The work was also supported by the Ministry of Education and Science of the Russian Federation, MEPhI Academic Excellence Project (contract N~02.a03.21.0005, 27.08.2013). 

\bibliographystyle{unsrt}

%\bibliography{bibt}

\end{document}